\documentclass[12pt]{article}
\usepackage{graphicx}
\usepackage{graphics}
\usepackage{fancybox}
\usepackage{amsmath,float,latexsym,psfrag,epsf,epsfig,amssymb,
chicago,rotating}
\usepackage{color}
\usepackage{url}

\usepackage{fullpage}

\newcommand{\pref}[1]{%
    \ref{#1} \ifnum\count0=\pageref{#1}\relax%
    \else (page \pageref{#1})\fi}

\newcommand{\eref}[1]{%
        \ref{#1}\ifnum\count0=\pageref{#1}\relax%
        \else {, p.\pageref{#1}}\fi}

\newcommand{\comment}[1]{}

\newenvironment{algorithm}{\vspace{5 mm}\sc}{\vspace{5 mm}}
\newlength{\labwidth}
\newcommand{\step}[1]{%
    \settowidth{\labwidth}{#1\ }%
    \par\noindent%
    \global\hangindent\labwidth {#1}%
    \hbox{ }%
    }%

\newcommand{\E}{{\mathbf{E}}}

\newcommand{\transpose}{{\mbox{\tiny T}}}
\newcommand{\diag}{{\mbox{diag}}}
\newcommand{\argmin}{{\mbox{arg min}}}
\newcommand{\argmax}{{\mbox{arg max}}}

\begin{document}

\begin{center}

{\Large \bfseries Estimating the model evidence: a review}
\vspace{5 mm}

{\large N. FRIEL\footnote{\texttt{nial.friel@ucd.ie}} and J. WYSE} \\
{\textit{School of Mathematical Sciences, University College Dublin, Ireland.}}
\vspace{5 mm}

\today

\vspace{5mm}

\end{center}

\bibliographystyle{mybib}

\begin{abstract}

\noindent The model evidence is a vital quantity in the comparison of statistical models under the Bayesian 
paradigm. This papers presents a review of commonly used methods. We outline some guidelines and offer some 
practical advice. The reviewed methods are compared for two examples; non-nested Gaussian linear regression and covariate subset selection in logistic regression.


\end{abstract}

\section{Introduction}

The advent of Markov chain Monte Carlo methods, since Geman and Geman \citeyear{gem:gem84} and especially 
Gelfand and Smith \citeyear{gel:smi90} have
led to what might be termed a Bayesian revolution. It is now routine to carry out a Bayesian analysis for
quite complex models with often high dimensional data. In particular, WinBUGS 
\shortcite{spie:tho:bes03} has allowed 
the practitioner access to many types of statistical models which are amenable to Gibbs sampling. This software 
has played a major role in the popularisation of Bayesian analysis to a very wide audience. 

More frequently, the practitioner is now not just interested in analysing data assumed from a single statistical model, but 
rather from the starting point where one assumes that there are a collection of models which could have 
plausibly generated the data. The usual objective is to understand the uncertainty associated with each statistical
model, or indeed to use this model uncertainty to aid prediction of, for example, future observations. 
The Bayesian paradigm offers a principled approach to the issue of model choice, through
examination of the model evidence, namely the probability of the data given the model. Suppose we are given data $y$ and 
assume there are a collection of competing models, $m_1,\dots,m_l$, each with associated parameters, $\theta_1,\dots,\theta_l$, 
respectively. Viewing the model indicators as parameters with prior distribution $\pi(m_k)$, the posterior distribution of 
interest is 
\[ \pi(\theta_k,m_k|y) \propto \pi(y|\theta_k,m_k)\pi(\theta_k|m_k)\pi(m_k) \]
where $ \pi(y|\theta_k,m_k)$ is the likelihood of the data under model $m_k$ with parameters $\theta_k$ and $\pi(\theta_k|m_k)$ is the prior on the parameters in model $m_k$.
The seminal paper by Green \citeyear{gre95} which introduced the reversible jump Markov chain Monte Carlo algorithm, allows
sampling of this joint model and parameter space and is itself a topic in this issue. This paper focuses on the complimentary
approach whereby one considers the posterior distribution conditional on model $m_k$, 
\[ \pi(\theta_k|y,m_k) \propto \pi(y|\theta_k,m_k)\pi(\theta_k|m_k). \]
The constant of proportionality for the un-normalised posterior distribution above is 
\[ \pi(y|m_k) = \int_{\theta_k} \pi(y|\theta_k,m_k)\pi(\theta_k|m_k)d\theta_k.\]
This is a vital quantity in Bayesian model choice and, somewhat confusingly is given different names in the
literature, the \textit{marginal likelihood}, \textit{integrated likelihood} or \textit{evidence}. The terminology, 
\textit{evidence} is the perhaps the most appealing and we will use this throughout this article. In general, for almost
all statistical models the evidence is analytically intractable, since this involves a high-dimensional integration over
a usually complicated and highly variable function. Computational tractability is rarely possible, and in such cases, relies
on conjugate priors. An interesting and recent example where conjugacy is used to this effect is in the paper by
Saban{\' e}s and Held \citeyear{bov:hel11}, in the context of generalized linear models. Additionally, the
evidence is very sensitive to the prior, so that small changes in the prior specification can lead to large changes in 
the evidence. 


This article is solely concerned with reviewing computational methods to estimate the model evidence while
the alternative approach to consider the posterior distribution of parameters and model indicators will be dealt with
in an article by Hastie and Green in this issue. 
Here we present some more recent developments in this area. This article is
aimed at readers not necessarily familiar with Bayesian methods. The intention is to provide a starting point from
which interested readers can delve further.

\section{Why do we need the evidence?}

Assuming that we could compute $\pi(y|m_k)$, then Bayes theorem can be used to combine the collection of evidence terms
for each model to form posterior model probabilities, 
\begin{equation}
\pi(m_k|y) = \frac{\pi(y|m_k)\pi(m_k)}{\sum_{j=1}^l \pi(y|m_j)\pi(m_j)}.
\label{eqn:post_mod}
\end{equation}
Moreover, if interest surrounded the comparison of two competing models, a natural quantity of interest is the Bayes
factor which is nothing more than the ratio of evidence terms for both models,
\[
 BF_{ij} = \frac{\pi(y|m_i)}{\pi(y|m_j)}.
\]
Indeed the ratio of posterior model probabilities can be expressed, using (\ref{eqn:post_mod}) as 
\[ \frac{\pi(m_i|y)}{\pi(m_j|y)} = \frac{\pi(y|m_i)}{\pi(y|m_j)}
    \times \frac{\pi(m_i)}{\pi(m_j)}. \]
The larger $BF_{ij}$ is, the greater the evidence in favour of $m_i$ compared
to $m_j$. Jeffreys \citeyear{jef61} presents a scale in which one can interpret the strength of evidence of one
model against another. The reader is also referred to Kass and Raftery \citeyear{kas:raf95} who present a comprehensive 
review of Bayes factors.

Bayesian model averaging \shortcite{hoe:mad:raf99} offers a very natural way to make predictions 
by averaging over all 
models, weighted proportional to their posterior model probabilities, thereby incorporating model uncertainty.
For example, if one were to make a prediction for a future observation (or indeed a collection of observations) $y^*$, 
then a natural quantity of interest is
\[
 \pi(y^*|y) = \sum_{k=1}^l \pi(y^*|m_k,y)\pi(m_k|y) .
\]
This is the average of the posterior distribution for $y^*$ under each model weighted
by the corresponding posterior model probabilities. 

\section{Review of methods to compute the model evidence}  \label{sec:methods}

Here we present, in chronological order, a short summary of various approaches to compute $\pi(y|m_k)$, the evidence for 
model $m_k$. In the remainder of this section we assume that we are estimating the evidence within a given model so that we suppress conditioning on the model $m_k$ and the subscript indexing the model on the parameters $\theta$. Subscripts on $\theta$ will be defined for the given context.

\subsection{Laplace's method}

An early and widely used method is Laplace's method \cite{tier:kad86}. This approach makes the
assumption that the posterior distribution can be adequately approximated by a normal distribution, for example if the
sample size is large enough. Obviously
this is a very strong assumption which might not often hold. Specifically, assume that $\pi(\theta|y)$ is highly 
peaked around the posterior mode $\tilde{\theta}$. 
Define
\[ l(\theta) = \log\{ \pi(y|\theta)\pi(\theta)\}. \]
Expanding $l(\theta)$ as a quadratic about $\tilde{\theta}$ and exponentiating leads to approximating
$\pi(y|\theta)\pi(\theta)$ as a Gaussian with mean $\tilde{\theta}$ and covariance 
 $\tilde{\Sigma} = (-D^2 l(\tilde{\theta}))^{-1}$, where $D^2 l(\tilde{\theta})$ is the Hessian matrix of 
second derivatives. Integrating this approximation yields
 \begin{equation}
  \pi(y) \approx (2\pi)^{d/2} |\tilde{\Sigma}|^{1/2} \pi(y|\tilde{\theta})\pi(\tilde{\theta}). \label{eq:LaplaceApproximation}
 \end{equation}
 
 \subsubsection{Running example: Gaussian model}
 
 As a means to demonstrating how the different approaches to estimating the evidence are applied in practice we adopt a simple running example which will be revisited after each approach is introduced.
We consider a sample of size $n$ independently distributed according to a $N(\mu,\tau^{-1})$ distribution. The priors on $\mu$ and $\tau$ are assumed $N(\xi,\nu^{-1})$ and $Ga(a_0/2,b_0/2)$ respectively. Here $\theta = (\mu,\tau)^{\transpose}$. Thus,
\[
\pi(y|\theta) = \left(\frac{\tau}{2\pi} \right)^{n/2}  \exp \left\{ -\frac{\tau}{2} \sum_{i=1}^{n} (y_i - \mu)^2\right\}
\]
and (ignoring unnecessary constants)
\[
l(\theta) = \left(\frac{n+a_0}{2} -1\right)\log \tau -\frac{\tau}{2}\left(\sum_{i=1}^n \left(y_i-\mu \right)^2  + b_0\right)- \frac{\nu}{2} (\mu-\xi)^2.
\]
The gradient is given by 
\[
\nabla l(\theta) = \left[\tau\left( \sum_{i=1}^n y_i - \mu \right) - \nu(\mu-\xi),  \left(\frac{n+a_0}{2} - 1\right)\frac{1}{\tau} - \frac{b_0}{2} - \frac{1}{2}\sum_{i=1}^n(y_i-\mu)^2\right]^{\transpose}
\]
and the Hessian is 
\[
D^2 l(\theta) = - \left(\begin{array}{cc} n\tau+\nu & n \mu - \sum_{i=1}^n y_i \\ 
 n \mu - \sum_{i=1}^n y_i &\left(\frac{n+a_0}{2} - 1\right)/\tau^2 \end{array} \right).
\]
It is possible to find the mode iteratively using a Newton method. Starting with a guess $\theta^{(0)}$ one iterates
\[
\theta^{(t+1)} = \theta^{(t)} - \gamma \left[ D^2l(\theta^{(t)})\right]^{-1} \nabla l(\theta^{(t)}), \qquad t = 0,1,\dots
\]
until some convergence criterion is satisfied ($\mbox{e.g.}$ $|l(\theta^{(t+1)}) - l(\theta^{(t)})| < 10^{-3}$). The parameter $\gamma$ in the iterations is a step-size which is usually chosen to be equal to 1. An estimate of $\pi(y)$ may then be obtained by taking $\tilde{\theta}$ as the last iterate and evaluating (\ref{eq:LaplaceApproximation}).

\subsection{Harmonic mean estimator}

The harmonic mean estimator \cite{new:raf94} is an easy to implement estimator of the evidence, 
based on draws from the target distribution. Suppose $\theta^{(1)},\dots,\theta^{(N)}$ is a sample generated from the posterior $\pi(\theta|y)$ using MCMC. Then after computing the likelihood of each $\theta^{(i)}$, $\pi(y|\theta^{(i)}), i = 1,\dots,N$, the evidence is estimated as the harmonic mean
of the likelihood,
\begin{equation}
 \pi(y) \approx 1/\left( \frac{1}{N} \sum_{i=1}^N \frac{1}{\pi(y|\theta^{(i)})} \right).
\label{eqn:harmonic}
\end{equation}
This estimator is easily shown to hold since it follows from the identity
\[
 \E\left\{ \frac{1}{\pi(y|\theta)}\right\} =
 \int \frac{\pi(y|\theta)\pi(\theta)}{\pi(y|\theta)\pi(y)}d\theta = 
 \frac{1}{\pi(y)}\int \pi(\theta)\;d\theta = \frac{1}{\pi(y)}.
\]
There is a severe downside to this estimator. Equation (\ref{eqn:harmonic}) is based solely on draws from the 
posterior. But the posterior is typically much more peaked than the prior, $\mbox{e.g.}$, when the posterior is insensitive 
to the prior. Hence in such situations, the harmonic mean estimator will not change much as the prior changes.
However it is well known that $\pi(y)$ is \textit{very sensitive} to changes in the prior. This drawback is very 
well documented, see \cite{rob:wra09}, for example. 

\subsubsection{Running example: Gaussian model}

This model may be sampled from using a Gibbs sampler. The sample $\theta^{(1)},\dots,\theta^{(N)}$ is drawn iteratively by drawing $\theta^{(i+1)} = (\mu^{(i+1)},\tau^{(i+1)})^{\transpose}$ conditioning on $\theta^{(i)}$:
\begin{equation}
\begin{array}{c}
\mu^{(i+1)} \sim N\left(\frac{\tau^{(i)} \sum\limits_{t=1}^n y_t + \nu \xi}{n \tau^{(i)} + \nu}, (n\tau^{(i)} + \nu)^{-1} \right) \\
\\
\tau^{(i+1)} \sim Ga \left( \frac{n+a_0}{2} , \frac{1}{2}\left[\sum\limits_{t=1}^n (y_t - \mu^{(i+1)})^2 + b_0 \right]\right).
\end{array}  \label{eq:GibbsSampler}
\end{equation}
The final sample $\theta^{(1)},\dots,\theta^{(N)}$ is obtained by throwing away some number of burn-in iterations at the beginning of the chain. 

To compute the harmonic mean estimator we calculate $\log \pi(y|\theta^{(i)}), i = 1,\dots,N$ and note that (\ref{eqn:harmonic}) may be rewritten
\[
\log \pi(y) \approx \log N - \log \left( \exp\{ - m \} \sum_{i=1}^N \exp \{ - \log \pi(y|\theta^{(i)}) + m \} \right)
\]
where $m = \max\{\log \pi(y|\theta^{(i)}),i = 1,\dots,N\}$. This approach to evaluating the harmonic mean should be more numerically stable than a direct implementation in cases where the log likelihoods are large negative numbers.

\subsubsection{A tractable example}

To illustrate the sensitivity of this estimator to changes in the prior consider the following simple example. Suppose that 
data $y_i\sim N(\mu,\tau^{-1})$, for 
$i=1,\dots,n$. Let's suppose that conjugate priors for $\mu$ and $\tau$ follow a $N(\mu_0,1/(\tau_0 \tau))$ and 
$Ga(a_0,b_0)$ distribution, respectively. This defines a so-called normal-gamma model 
\cite{ber:smi94}, for which the full-conditional distributions are
\[
 \mu|\tau \sim N(\mu_n,\tau_n^{-1}) \;\;\mbox{and}\;\; \tau|\mu \sim Ga(a_n,b_n),
\]
where
\[
 \mu_n = \frac{\tau_0\mu_0+n\bar{y}}{\tau_0+n}; \;\; \tau_n = \tau_0+n;\;\; a_n = a_0+n/2; \;\; 
 b_n = b_0 + \frac{1}{2}\sum_{i=1}^n (y_i-\bar{y})^2 + \frac{\tau_0 n(\bar{y}-\mu_0)^2}{2(\tau_0+n)}.
\]
This model is also one for which the evidence can be computed analytically,
\[
\pi(y) = \frac{\Gamma(a_n)}{\Gamma(a_0)} \frac{b_0^{a_0}}{b_n^{a_n}}\left( \frac{\tau_0}{\tau_n} \right)^{1/2}(2\pi)^{-n/2}.
\]

Here we simulated $n=100$ observations, with $\mu=0$ and $\tau=1$. Prior parameters for $a_0$, $b_0$ and $\mu_0$ were set
to $0.001$, $0.001$ and $0$, respectively. Our interest was to see how the harmonic mean estimator performed for varying
values of $\tau_0$. In Table~\ref{tab:harmonic_mean}, the harmonic mean estimate of the logarithm of the evidence, 
$\log \hat{\pi}(y)$, based on $10^6$ iterations of a Gibbs sampler, is presented for varying values of $\tau_0$. 
Clearly, these estimates do not vary much with $\tau_0$, in contrast to the true value of $\log \pi(y)$ which varies 
as the prior for $\mu$ changes.
\begin{table}[htp]
 \centering
 \begin{tabular}{c|cccc}
   $\tau_0$ & $0.0001$ & $0.01$ & $0.1$ & $1$ \\
   \hline
   $\log \pi(y)$ & $-160.54$ & $-158.24$ & $-157.09$ & $-155.95$ \\
   $\log \hat{\pi}(y)$ & $-148.33$ & $-148.45$ & $-148.59$ & $-148.85$
 \end{tabular}
\caption{The true value of $\log \pi(y)$ is presented for various values of $\tau_0$, together
with estimates of the logarithm of $\pi(y)$, based on the harmonic mean estimator.}
\label{tab:harmonic_mean}
\end{table}

As above, the explanation for the poor performance of the harmonic mean estimator lies in the fact
that the posterior distribution does not change very much, as the prior distribution for $\mu$ becomes more
diffuse, from $\tau_0=1$ through to $\tau_0=0.0001$. Therefore, samples from the four different posterior distributions
in this example will be very similar, as reflected in the similar harmonic mean estimates. However, the evidence does
depend on how diffuse the prior for $\mu$ is, since it involves an integration over $\mu$ (and $\tau$). 

\subsection{Chib's method}

Chib \citeyear{chib95} presented a generic method which can be applied to output 
from the Gibbs sampler. This estimator follows from re-arranging Bayes' formula to yield
\[ \pi(y) = \frac{\pi(y|\theta)\pi(\theta)}
   {\pi(\theta|y)}.
\]
Using this identity one could estimate $\log \pi(y)$ as
\begin{equation} 
 \log \pi(y) = \log \pi(y|\theta^*) +
   \log \pi(\theta^*) - \log \hat{\pi}(\theta^*|y)
\label{eqn:chib}
\end{equation}
where, for example, $\hat{\pi}(\theta^*|y)$ is an estimate of the posterior density 
at a point $\theta^*$ of high posterior probability, although the equality holds for any $\theta^*$. 
Examining the right hand side of (\ref{eqn:chib}), both the prior and likelihood terms, $\pi(\theta^*)$
and $\pi(y|\theta^*)$, respectively, can usually be evaluated in closed form, for most statistical models.
The estimation of the posterior probability, $\hat{\pi}(\theta^*|y)$, is troublesome. The important
contribution of Chib was to realise that $\pi(\theta^*|y)$ can be estimated by a Monte Carlo average
based on draws from the Gibbs sampler. 

Suppose the vector $\theta$ can be partitioned as $(\theta_1,\theta_2,\theta_3)$,
where the full-conditional distribution of each $\theta_i$ is of standard form. The law of total
probability allows one to write
\[\pi(\theta^*|y) = \pi(\theta^*_1|\theta^*_2,\theta^*_3,y)
\pi(\theta^*_2|\theta^*_3,y) \pi(\theta^*_3|y) \]
Gibbs sampling can be used to estimate each factor on the left hand side, 
\begin{eqnarray*}
 \hat{\pi}(\theta^*_3|y) &=& \frac{1}{N}\sum_{i=1}^N \pi(\theta^*_3|\theta_1^{(i)},\theta^{(i)}_2,y).\\
 \hat{\pi}(\theta^*_2|\theta^*_3,y) &=& \frac{1}{N}\sum_{i=1}^N \pi(\theta^*_2|\theta_1^{(i)},\theta^*_3,y).
\end{eqnarray*}
Clearly, this approach can be extended to high-dimensional parameter spaces. Chib's method can
obviously be used in many situations, given the wide applicability of the Gibbs sampler for a very wide 
class of statistical models. Certainly, this is the case -- it is clear from the number of citations which
this article has gained that this approach is popular among practitioners. In terms of computational implementation, 
a requirement is that draws from various full-conditional distributions need to be stored, although this is
rarely prohibitive. Note also that this method has been extended \cite{chib:jel01}
so that the evidence can be estimated based on output from a Metropolis-Hasting sampler. 
Finally, it should be mentioned that difficulties can arise when this method is applied to mixture models, hidden
Markov models and other models which give rise to label switching and parameter non-identifiability. The difficultly can arise because 
for a finite mixture model with $k$ components, for example, it can take a long time, in practice, for a Gibbs
sampler to visit all the unique $k!$ modes. See 
{\texttt{www.cs.utoronto.ca/\%7Eradford/ftp/chib-letter.pdf}} for a more detailed discussion of this issue.

\subsubsection{Running example: Gaussian model}

The parameter $\theta$ may be partitioned in this case into $(\mu,\tau)$. Draw a sample $\theta^{(1)},\dots,\theta^{(N)}$ using the Gibbs sampling scheme (\ref{eq:GibbsSampler}). We will first evaluate $\hat{\pi}(\tau^*|y)$. This is done by computing
\[
\hat{\pi}(\tau|y) = \frac{1}{N}\sum_{i=1}^N Ga\left(\tau\left|\frac{n+a_0}{2},\frac{1}{2}\left[ \sum_{t=1}^n (y_t - \mu^{(i)})^2 +b_0 \right]\right.\right)
\]
where $Ga(x|a,b)$ represents the gamma density function with shape $a$ and rate $b$ evaluated at $x$. This should be maximized over the range of $\tau$ for a good approximation to the posterior ordinate. For example, one could compute $\hat{\pi}(\tau^{(j)}|y)$ for each $\tau^{(j)}$ in the sample, and then take $\tau^* = \argmax_{\tau \in \{\tau^{(1)},\dots,\tau^{(N)}\}} \hat{\pi}(\tau|y)$.

Equipped with $\tau^*$ we can compute the mode of the full conditional of $\mu|\tau^*$. This is more straightforward since $\mu|\tau^*$ is normally distributed with mode at 
$
\mu^* = \left(\tau^* \sum_{t=1}^n y_t + \nu \xi\right)/(n \tau^* + \nu)
$
so that $\hat{\pi}(\mu^*|\tau^*,y) = (n\tau^* + \nu)^{1/2} / \sqrt{2\pi}$. Denoting $\theta^*  = (\mu^*,\tau^*)^{\transpose}$,
\[
\log \hat{\pi}(\theta^*|y) = \frac{1}{2} \log(n\tau^* + \nu) - \frac{1}{2} \log (2\pi) + \log\hat{\pi}(\tau^*|y)
\]
which can be used to obtain (\ref{eqn:chib}) with $\theta^*$.

\subsection{Annealed importance sampling}

Importance sampling plays a vital role in Monte Carlo sampling. Suppose that the target distribution of interest
is $\pi(x)$, which is not amenable to direct sampling, but that $\hat{\pi}(x)$ is an approximation of it, from which
one can sample easily referred to as the importance function. Here $x$ is used to generically denote some random variable and $\pi(\cdot)$ its density, most likely up to an unknown constant of proportionality. Then draws $x^{(1)},\dots,x^{(N)}$ from $\hat{\pi}(x)$ can be used as an importance 
sampling estimate of the expectation of some $\pi$-integrable function $a(\cdot)$ since
\[
 \E_{\pi} a(x) = \int a(x) \pi(x) \,\mbox{d} x =  \int a(x) \frac{\pi(x)}{\hat{\pi}(x)} \hat{\pi}(x) \,\mbox{d} x = 
\E_{\hat{\pi}} w(x)a(x) \approx \frac{1}{N}\sum_{i=1}^N w(x^{(i)})a(x^{(i)}),
\]
where the importance weights are defined as
\[
 w(x)=\frac{\pi(x)}{\hat{\pi}(x)}
\]
once $\pi(\cdot)$ and $\hat{\pi}(\cdot)$ are normalized. 

When one or both of $\pi(\cdot)$ and $\hat{\pi}(\cdot)$ are not normalized, the ratio of their normalizing constants $z_{\pi}/z_{\hat{\pi}}$ (with $z_{\pi}=\int_{x}\pi(x)\,\mbox{d}x$ and $z_{\hat{\pi}}=\int_{x} \hat{\pi}(x)\,\mbox{d}x$) may be approximated by noting
\[
\frac{z_{\pi}}{z_{\hat{\pi}}} = \frac{1}{z_{\hat{\pi}}} \int \pi(x) \, \mbox{d} x = \frac{1}{z_{\hat{\pi}}} \int\frac{ \pi(x)}{\hat{\pi}(x)} \hat{\pi}(x) \, \mbox{d} x = \int w(x) \frac{ \hat{\pi}(x)}{z_{\hat{\pi}}} \, \mbox{d} x = \E_{\hat{\pi}} w(x)
\]
so that 
\begin{equation}
\frac{1}{N}\sum_{i=1}^N w(x^{(i)}) \rightarrow \frac{z_{\pi}}{z_{\hat{\pi}}} \label{eqn:imp_samp}
\end{equation}
as $n \rightarrow \infty$. Then in the case of un-normalized densities the importance sampling estimate above is given by,
\[
\E_{\pi} a(x) = \frac{\E_{\hat{\pi}} a(x)}{z_{\pi}/z_{\hat{\pi}}} \approx 
\frac{\frac{1}{N}\sum_{i=1}^N a(x^{(i)}) w(x^{(i)})}{\frac{1}{N}\sum_{i=1}^N w(x^{(i)})}.
\]


An important issue is choice of the importance function, which generally needs to have heavier tails than the
target distribution. Annealed importance sampling (AIS), developed by Neal \citeyear{neal01} is a very 
clever algorithm which, broadly speaking, uses a tempering mechanism to adaptively define an importance sampling 
function to approximate the target posterior distribution. This is done by defining 
\[
 \pi_{t_j}(\theta|y) = \pi(\theta)^{1-t_j} \pi(\theta|y)^{t_j}, 
\;\mbox{where}\; 1=t_0>\dots >t_m=0. 
\]
where $\pi(\theta|y)^{t_j}$ is an un-normalised density composed by raising the (un-normalised) posterior to power $t_j$. Then 
$\{\pi_{t_0}, \dots, \pi_{t_m}\}$ defines a sequence of distributions, transitioning
from the prior distribution to the posterior distribution. The innovative aspect of this algorithm is to use a Markov
transition kernel, for example, a Metropolis-Hastings kernel, to allow one to draw from an importance function which can be 
used to approximate $\pi(\theta|y)$, the target distribution.

Let $T_j$ denote a Markov transition kernel with invariant $\pi_{t_j}$, for example, one based on Metropolis-Hastings
updates. The AIS algorithm can now be sketched as follows.

\begin{algorithm}
 \step{} for $i=1,\dots,N$ do: 
 \step{} \indent sample $\theta_{m-1}$ from $\pi_{t_m}$. 
 \step{} \indent sample $\theta_{m-2}$ from $\theta_{m-1}$ using $T_{m-1}$ 
 \step{} \indent $\cdots$ 
 \step{} \indent sample $\theta_0$ from $\theta_1$ using $T_1$
 \step{} \indent set
\[ 
  \theta^{(i)}=\theta_0\;\; \mbox{and}\;\;
  w(\theta^{(i)}) = \frac{\pi_{t_{m-1}}(\theta_{{m-1}}|y)}{\pi_{t_m}(\theta_{m-1}|y)}\frac{\pi_{t_{m-2}}(\theta_{m-2}|y)}
  {\pi_{t_{m-1}}(\theta_{m-2}|y)}\dots \frac{\pi_{t_{0}}(\theta_{0}|y)}{\pi_{t_1}(\theta_{0}|y)}.
\]
 \step{} end
\end{algorithm}

As is the case for standard importance sampling, AIS yields an independent sample $\{\theta_i: \; i=1,\dots,N\}$ 
from the target distribution and this is a major strength of this approach. But also, by analogy to (\ref{eqn:imp_samp}), 
it is possible to show that the average of the importance weights yield an estimator of the ratio of the normalising 
constants of the target distribution and the prior distribution, that is, an estimator of the evidence. 
\[ \pi(y) \approx \frac{1}{N}\sum_{i=1}^N   w(\theta^{(i)}). \]

AIS is very well suited to complicated posterior distributions which present difficulties for standard MCMC algorithms, 
in particular, the tempered aspect of the algorithm
should help in situations where the posterior distribution is multi-modal. Note that from an implementation
point of view, there are some key aspects to AIS. The choice of the transition kernel corresponding to the 
distribution at temperature $t_j$ is vital, and in practice one may need a few full sweeps of the parameters 
in order that the sequence $\theta_{m-1},\dots,\theta_1$ yields a representative draw $\theta^{(i)}$ from the
target distribution. In common with tempering algorithms, the choice of the temperature ladder $t_m,\dots,t_0$
is vital. There has been some work in this direction for the related tempered transition algorithm 
\cite{neal96}, see \shortcite{beh:fri:hur11}. AIS is a popular method 
and is highly cited in the literature. It appears that it is primarily cited in the machine learning literature, and 
there is certainly much scope for it to apply in mainstream statistical settings. 

\subsubsection{Running example: Gaussian model}

We first define the modified posteriors $\pi_{t_j}(\theta|y)$ which in this case are 
\begin{equation}
\pi_{t_j}(\theta|y) = \pi(\theta)^{1-t_j} \pi(\theta|y)^{t_j} \propto  \pi(\theta)^{1-t_j} \left\{\pi(y|\theta)\pi(\theta) \right\}^{t_j} = \pi(\theta) \pi(y|\theta)^{t_j}. \label{eq:AISPosterior}
\end{equation}
We must choose the transition kernel $T_j$ with invariant $\pi_{t_j}$. Here it is possible to choose a Gibbs kernel, since the full conditionals of $\mu$ and $\tau$ at temperature $t_j$ are
\begin{equation}
\begin{array}{c}
\mu|\tau,t_j \sim  N\left( \frac{t_j n \sum\limits_{t=1}^n y_t + \nu \xi}{t_j n \tau + \nu}, (t_j n \tau + \nu)^{-1} \right)\\
\tau|\mu,t_j  \sim  Ga\left(\frac{t_j n + a_0}{2}, \frac{1}{2}\left[t_j \sum\limits_{t=1}^n (y_t-\mu)^2 + b_0\right] \right).
\end{array} \label{eqn:TemperedPosterior}
\end{equation}

In implementing AIS, we would draw $\theta_{m-1} = (\mu_{m-1},\tau_{m-1})^{\transpose}$ from the joint prior $\pi(\mu,\tau) = \pi(\mu)\pi(\tau)$, then $\theta_{m-k}$ would be sampled from $\theta_{m-k+1}$ using $T_{m-k+1}$ , $k = 2,\dots,m$. It will be necessary in practice to apply $T_{m-k+1}$ a number of times to ensure that $\theta_{m-k}$ is approximately distributed according to $\pi_{t_{m-k+1}}$.

\subsection{Nested sampling}

Nested sampling \cite{skilling06} is a generic algorithm for computing the evidence. The 
evidence may be viewed as the expectation, with respect to the prior, of the likelihood. As such, the evidence 
can be expressed as
\[
 \pi(y) = \int \pi(y|\theta) \pi(\theta)\; \mbox{d}\theta = \int \pi(y|\theta)\; \mbox{d}X,
\]
where $\mbox{d}X = \pi(\theta)\;\mbox{d}\theta$ is an element of prior mass. 

Define
\[
 X(\lambda) = \int_{\pi(y|\theta)>\lambda} \pi(\theta)\;\mbox{d}\theta
\]
as a cumulant prior mass. Writing the inverse function as $\pi(y|X)$, that is, $\pi(y|X(\lambda))=\lambda$ 
allows the evidence to be expressed as a $1-$dimensional integral:
\[
 \pi(y) = \int_0^1 \pi(y|X)\; \mbox{d}X.
\]
Suppose that we know how to evaluate the likelihood as $l_i = \pi(y|X_i)$ at a right-to-left sequence of $I$ points $0<X_I < \dots < X_2 < X_1 < 1$. Then any convenient quadrature method would estimate the evidence as a weighted sum $\sum_{i = 1}^{I} w_i l_i$. Since $\pi(y|X)$ is non-increasing, it will be bounded below by any value evaluated at a larger value of $X$. Hence $w_i = X_i - X_{i+1}$ with $X_{I+1} = 0$ gives a lower bound for estimating $\pi(y)$.

The main computational burden of nested sampling is the requirement to sample $\theta$ from the prior 
subject to the constraint that $\pi(y|\theta)>l$, for value $l$. This is roughly similar to the computational
cost of slice sampling \cite{neal03}. The evidence is estimated by sorting draws from the prior according 
to their likelihood.
\[
 \pi(y)  = \sum_{i=1}^{I} (X_i - X_{i+1})\pi(y|\theta^{(i)}).
\]
The nested sampling algorithm may be sketched as follows.

\begin{algorithm}
 \step{} Sample $\theta^{(1)},\dots,\theta^{(N)}$ from the prior and initialize $\hat{\pi}(y) = 0$ 
 \step{} For $i=1,\dots,I$ do: \\
 \step{} \indent Find point $\theta_i$ with the smallest likelihood, $l_i$, among the $N$ current 
 \step{} \indent $\theta^{(j)}$'s, $j=1,\dots,N$. 
 \step{} \indent Set $X_i = \exp\left\{-i/N\right\}$ and $w_i = X_{i-1}- X_i$.
 \step{} \indent Increment $\hat{\pi}(y)$ by $l_i w_i$.
 \step{} \indent Replace $\theta_i$ with a point sampled from the prior subject to $\pi(y|\theta)>l_i$.
 \step{} end
\end{algorithm}

\noindent Note that the deterministic step $X_i = \exp\{-i/N\}$ could be replaced by a stochastic step. The 
algorithm may be terminated after a given number of iterations, when new contributions to $\hat{\pi}(y)$ are 
below some small fraction of the current value as suggested by Chopin and Robert \citeyear{chop:rob10}, or 
using one of the criteria outlined in \cite{skilling06}.

Nested sampling is a popular algorithm in the astronomy and related literatures. Yet it appears
not to be widely used in mainstream statistics. Arguably, this is due to the computational overhead
required to implement this algorithm. Note that Chopin and Robert \citeyear{chop:rob10} illustrate,
for the case where levels $X_i$ follow a deterministic scheme, that the convergence rate is $O(N^{1/2})$. 
Their results hold where importance sampling is used to carry out the constrained prior sampling, and
they note that similar results for an MCMC implementation of nested sampling are unknown. Finally
Chopin and Robert \citeyear{chop:rob10} illustrate that the algorithm performed well for a mixture
model and a probit model, but that further experimentation is needed to determine situations where
it may or may not work well. 

\subsubsection{Running example: Gaussian model}

We would begin here by simulating $\theta^{(i)} = (\mu^{(i)},\tau^{(i)})^{\transpose}, i =1,\dots,N$ from the prior and computing $\log \pi(y|\theta^{(i)})$ for each. Then 
\[
\theta_1 = \argmin _{\theta \in \{\theta^{(1)},\dots,\theta^{(N)}\}} \log \pi(y|\theta)
\]
and $l_1 = \pi(y|\theta_1)$. 

We build up an approximation $a$ to $\log \pi(y)$ by initializing this to the negative of the largest floating point number available in the machine on which computations are being carried out. Then $a$ is incremented by $l_1 + \log(1 - \exp\{-1/N\})$. We continue sampling from the prior until we obtain a $\theta$ such that $\pi(y|\theta)>l_1$, and replace $\theta_1$ from the original sample from the prior with $\theta$ and leaving all other $\theta^{(i)}$ in the sample unchanged. The process is then repeated giving
\[
\theta_2 = \argmin _{\theta \in \{\theta^{(1)},\dots,\theta^{(N)}\}} \log \pi(y|\theta)
\]
and $l_2 = \pi(y|\theta_2)$, and $a$ is incremented by $l_2 + \log(\exp\{-1/N\} - \exp\{-2/N\})$. This is iterated until a new contribution to $a$ is less than $-8 \log 10 + a$, corresponding to a new contribution to the sum being less than $10^{-8}$ times the current value. 

An alternative to sampling from the prior until the constraint $\pi(y|\theta)>l_i$ is satisfied, would be to choose one of the remaining sample $\theta$ at random (which is already known to have $\pi(y|\theta)>l_i$) and move this a number of times subject to the constraint using a Metropolis-Hastings algorithm with target distribution $\pi(\theta)$. An example would be
\begin{algorithm}
\step{} Select $\theta$ at random from the remaining sample not minimizing the likelihood.
\step{} Draw $\theta^{'} \sim q(\cdot|\theta)$ from some proposal $q(\cdot)$ depending on $\theta$.
\step{} If $\pi(y|\theta^{'}) < l_i$, reject the proposal, otherwise accept with probability \[\min\left(1,\frac{\pi(\theta^{'})q(\theta|\theta^{'})}{\pi(\theta)q(\theta^{'}|\theta)} \right)\]
\step{} Repeat this process a given number of times until the dependency on $\theta$ is forgotten.
\end{algorithm}

When we terminate the algorithm we add an end correction to $a$ which should be negligible. See 
\cite{skilling06} and the pseudocode in the appendix there for details.

\subsection{Power posteriors}

In statistical
physics, there is a large body of work concerned with methods for estimating normalising constants, or partition 
functions of statistical models. A notable example is the Ising model. The method of thermodynamic integration 
was developed in the statistical physics 
literature to handle such problems, and an excellent review of this approach, and other related methods, from a
statistics perspective can be found in \cite{gel:men98}. 
Friel and Pettitt \citeyear{fri:pet08} explored how
thermodynamic integration could be used for the specific instance where the normalising constant is the 
evidence arising from an un-normalised posterior distribution. Specifically they consider what they term the power posterior,
\[ \pi(\theta|y,t) \propto \{\pi(y|\theta)\}^t
   \pi(\theta), \]
where $t\in [0,1]$. 
By construction the normalising constant of the power posterior is
\[ z(y|t) = \int_{\theta} \{\pi(y|\theta)\}^t
\pi(\theta)\;\mbox{d}\theta, \]
where $z(y|t=1)$ is the model evidence and $z(y|t=0)$ is the integral of the prior for $\theta$, 
which equals $1$.

The model evidence follows the identity:
\begin{equation}
  \log \pi(y) = \log\left\{ \frac{z(y|t=1)}{z(y|t=0)}
  \right\} = \int_0^1 \E_{\theta|t}
  \log{\pi(y|\theta)}\mbox{d}t. \label{eq:evid}
\end{equation}
To prove this one can show that the gradient of the log normalising constant can be expressed as
the expected posterior deviance. 
\begin{eqnarray}
  \frac{\mbox{d}}{\mbox{d}t} \log(z(y|t)) &=& \frac{1}{z(y|t)}
  z^{\prime}(y|t) \nonumber \\
  &=& \frac{1}{z(y|t)}\int
  \frac{\mbox{d}}{\mbox{d}t}\log(\pi(y|\theta))^t \pi(\theta)\,\mbox{d}\theta
  \nonumber \\
  &=& \int \log(\pi(y|\theta)) \frac{\pi(y|\theta)^t
  \pi(\theta) }{z(y|t)}\, \mbox{d}\theta \nonumber \\
  &=& \E_{\theta|t} \log(\pi(y|\theta)). 
 \label{eqn:power_post}
\end{eqnarray}
Integrating (\ref{eqn:power_post}) with respect to $t$ yields (\ref{eq:evid}).
In practice an estimator based on (\ref{eq:evid}) is formulated by discretising $t\in{[0,1]}$, 
$0=t_0<t_1,\dots,t_m=1$. For each $t_j$, a sample from $\pi(\theta|y,t_j)$ can be used
to estimate $E_j = \E_{\theta|t_j} \log{\pi(y|\theta)}$. Finally, a trapezoidal rule can be used
to approximate  
\begin{equation}
\log \pi(y) \approx \sum_{j=1}^m (t_{j}-t_{j-1}) \left( \frac{E_{j-1}+E_j}{2} \right) \label{eqn:quadrature}
\end{equation}
 
Discretising the temperature introduces an error. The other source of error is the Monte Carlo error arising
from approximating $\E_{\theta|t_j} \log{\pi(y|\theta)}$. 
Calderhead and Girolami \citeyear{cal:gir09} have shown that the discretisation error depends on the 
Kullback-Liebler distance between
$\pi_{t_j}$ and $\pi_{t_{j+1}}$, for $j=0,\dots,m-1$. Thus for a fixed number of temperatures, 
the optimal positioning of the $\{t_j\}$, in terms of minimising the error due to the temperature discretisation, should 
be so that the Kullback-Liebler distance between successive tempered distributions is minimised.

The power posterior approach is a generic method for estimating the evidence, and is relatively easy
to implement. It is also often possible to implement it in WinBUGS. However, an immediate difficulty
with this approach is that of choosing the temperature schedule, and this could be viewed as a weakness,
although Behrens {\it et al}. \citeyear{beh:fri:hur11} offer some possibilities in this direction.

\subsubsection{Running example: Gaussian model}

Here we may sample within a given temperature using the Gibbs sampler in (\ref{eqn:TemperedPosterior}). After generating $N$ samples $\theta^{(1)},\dots,\theta^{(N)}$ within temperature $t_j$ we estimate
\[
E_j \approx \sum_{i=1}^N \log \pi(y|\theta^{(i)}).
\]
To get starting values for the chain at temperature $t_{j+1}$ we use the average of the sampled values at $t_j$,
\[
\begin{array}{cc}
\mu_{\mbox{\tiny s}}^j = \frac{1}{N}\sum_{i=1}^N \mu^{(i)} \qquad \tau_{\mbox{\tiny s}}^j = \frac{1}{N}\sum_{i=1}^N \tau^{(i)} .
\end{array}
\]
We sample for all temperatures $t_j$ and then estimate the log evidence using (\ref{eqn:quadrature}).

\subsection{Other approaches}

This article has focused mostly on evidence estimation methods based on MCMC sampling. However other
approaches are also possible. Indeed evidence estimation is also possible using sequential Monte Carlo
(SMC) methods, which is not surprising given that it is close in spirit to tempering methods. See 
Section $4$ of \shortcite{del:dou:jas06}, where the evidence is estimated for a mixture model 
using SMC and AIS. For this example the authors conclude that SMC does not give superior performance to
AIS. Note also that the authors argue that the issue of optimal temperature schedule (or path) in sequencing
from, for example, prior to posterior is an important, but very challenging problem. This issue is clearly also
of relevance to AIS and to the power posterior approach. Variational Bayes methods have been used to estimate 
the evidence for Ising models \cite{parise:welling06}, for example. 
Finally, we alert the reader to the fact that the integrated nested Laplace approximation (INLA) framework 
\shortcite{rue:mar:cho09} could potentially be useful as a means to provide an estimate of the evidence for the 
class of statistical models which can be represented as Gaussian Markov field models. This includes a wide 
class of models from auto-regressive time series models to stochastic volatility models. INLA software 
(\url{www.r-inla.org/}) gives the possibility to analyse such models providing an estimate of the model evidence. 
See for example, \shortcite{wyse:fri11} in the context of change-point models with dependence 
between observations. 

\section{Some examples}

Here we provide a brief numeric comparison of the methods reviewed in Section~\ref{sec:methods} on two full examples. The first of these compares two simple Gaussian linear non-nested regression models. The prior assumptions here lead to analytically tractable models and thus exact evaluation of the evidence and Bayes factor between the models, giving a benchmark for comparison of the methods in Section~\ref{sec:methods}. The second example applies each of the methods to computing the evidence for competing logistic regression models. In this case there is no analytic tractability and the evidence must always be estimated via Monte Carlo sampling or other approximate techniques.

\subsection{Non-nested linear regression models}

This data describes the maximum compression strength parallel to the grain $y_i$, density $x_i$ and density adjusted for 
resin content $z_i$ for $n = 42$ specimens of {\it radiata} pine. This data originates from \cite{williams59}. 
It is wished to determine whether the density or resin-adjusted density is a better predictor of compression strength parallel 
to the grain. With this in mind, two Gaussian linear regression models are considered;
\[
\begin{array}{lccc}
\mbox{Model 1: } & y_i = \alpha + \beta(x_i - \bar{x}) + \epsilon_i, & \epsilon_i \sim N(0,\tau^{-1}), &i = 1,\dots,n,\\
\mbox{Model 2:} & y_i = \gamma + \delta(z_i - \bar{z}) + \eta_i, & \eta_i \sim N(0,\lambda^{-1}), & i = 1,\dots,n.\\
\end{array}
\]
The priors assumed for the line parameters $(\alpha,\beta)^{\transpose}$ and $(\gamma,\delta)^{\transpose}$ had 
mean $(3000,185)^{\transpose}$ with precision (inverse variance-covariance) $\tau Q_0$ and $\lambda Q_0$ respectively 
where $Q_0 = \diag(r_0,s_0)$. The values of $r_0$ and $s_0$ were taken to be 0.06 and 6. A gamma prior with shape 
$a_0 = 6$ and rate $b_0 = 4\times300^2$ was taken for both $\tau$ and $\lambda$. These prior assumptions give rough 
equivalence with the priors assumed for this data in other analyses. See for example \cite{fri:pet08}.


It is possible to compute the exact marginal likelihood for both of these models due to the prior assumption that the precision 
on the mean of the regression line parameters is proportional to the error precision. For example, the marginal likelihood of 
Model 1 is given by
\[
\pi(y) =  \pi^{-n/2} b_0^{a_0/2}\frac{\Gamma\left\{(n+a_0)/2 \right\}}{\Gamma\left\{ a_0/2\right\}}\frac{|Q_0|^{1/2}}{|M|^{1/2}} \left(y^{\transpose} R y + b_0 \right)^{-(n+a_0)/2}
\]
where $y  = (y_1,\dots,y_n)^{\transpose}$, $M = X^{\transpose} X + Q_0$ and $R  = I - X M^{-1} X^{\transpose}$ with the 
$i^{\mbox{th}}$ row of $X$ equal to $(1\,\,\, x_i)$ and $I$ is the $2\times 2$ identity matrix.

\begin{figure}
\begin{center}
\[
\begin{array}{ccc}
\includegraphics[width=50mm]{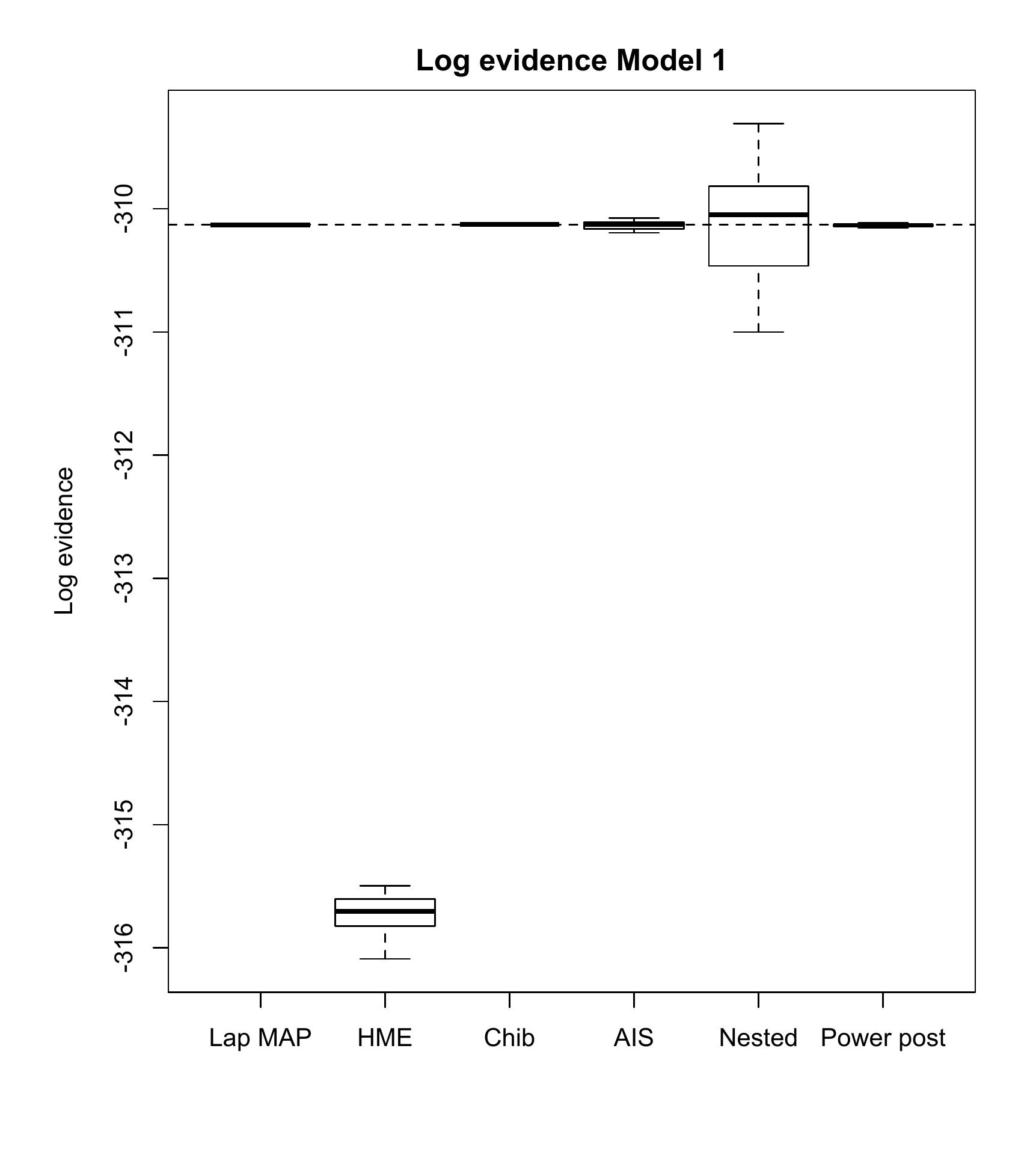}& \includegraphics[width=50mm]{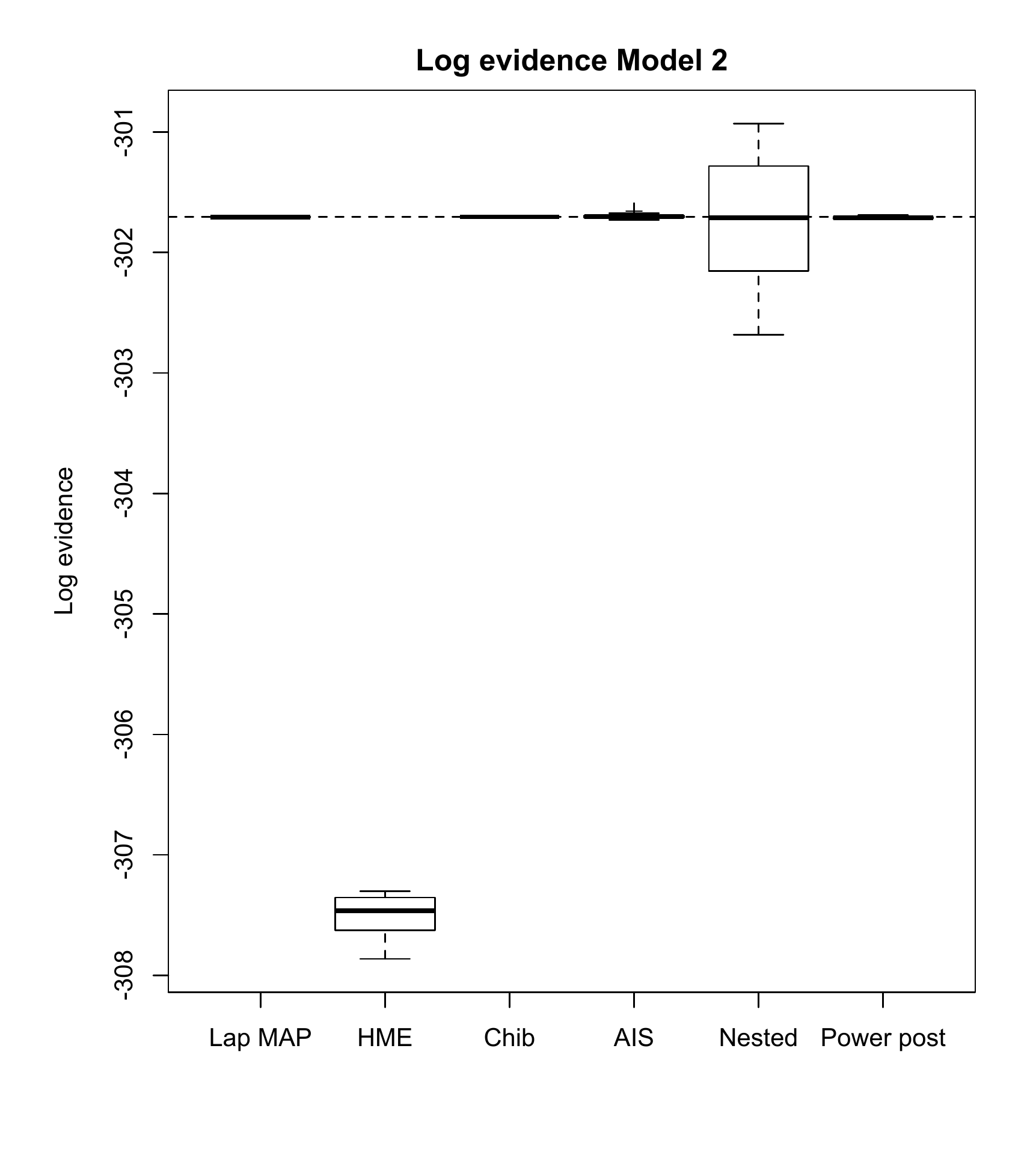} & \includegraphics[width=50mm]{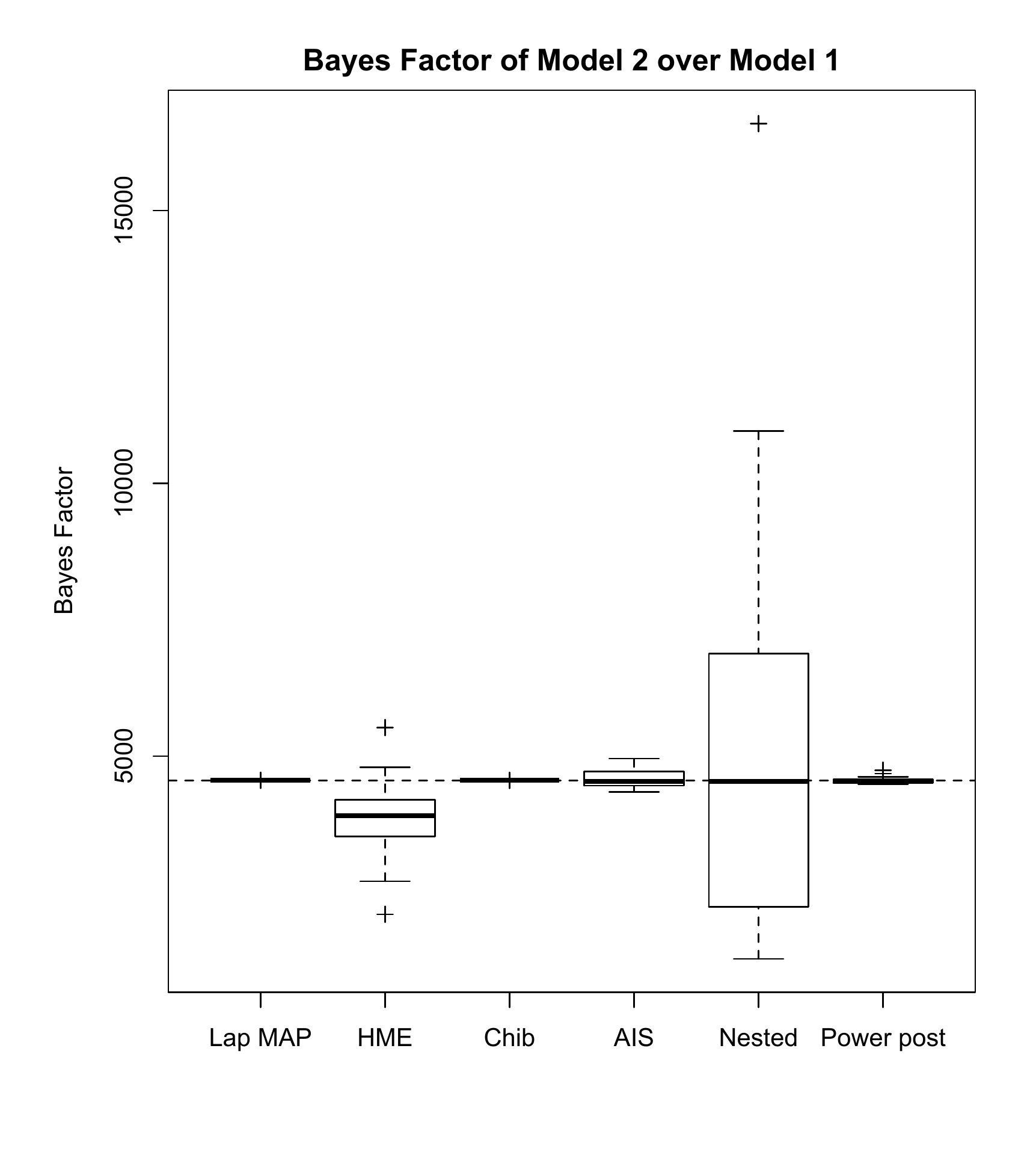}
\end{array}
\]
\end{center}
\caption{Boxplots of evidence models 1 \& 2 (left and middle panel) and computed Bayes factors (right panel) over eighteen runs of the algorithms of Section~\ref{sec:methods}. The true value is shown by the dotted line.} \label{fig:boxplot}
\end{figure}

The exact value of the Bayes factor of Model 2 over Model 1 is given in Table~\ref{tab:comparison} along with the Laplace 
approximation to this and mean and standard deviation of the estimate computed over eighteen runs of the Monte Carlo and nested sampling algorithms. 
For the Laplace approximation, the mode was located by using a standard Newton scheme. Out of interest we also computed the 
Laplace approximation at the MAP (maximum {\it a posteriori}) value of a simulation from the posterior. A series of boxplots 
illustrating the estimated log evidences and corresponding Bayes Factors computed over the eighteen runs is given in 
Figure~\ref{fig:boxplot}.

To make the each of the approaches for computing the evidence as comparable as possible, we tried to make each of the algorithms 
equivalent in terms of the number of iterations. For the Laplace approximation at the MAP and the harmonic mean estimator, 
results were based on a Gibbs sampling run of 505,000, taking 20\% of this as a burn-in. For Chib's method, a Gibbs run sampling 
all the model parameters was run for 55,000 burn-in iterations with a subsequent 150,000 iterations before fixing the precision 
at it's modal value. Following this a run of 150,000 iterations for each of the regression line parameters in turn was carried 
out, fixing the previous line parameters at their modal value. For annealed importance sampling 1000 samples were generated from 
the posterior taking a geometric temperature ladder $t_i = (1-i/100)^5, i = 0,\dots,100$. Within each step of the ladder 
we sample five times from each tempered posterior to ensure a representative sample. Power posteriors used an identical 
(reversed) temperature ladder  $t_i = (i/100)^5, i = 0,\dots,100$ to annealed importance sampling. Within each step of the ladder 
we ran the Gibbs sampling scheme for 5000 iterations using 20\% of this as a burn-in. Finally for nested sampling we terminated the 
algorithm when a new contribution to the nested sampling approximation of the evidence was less than $10^{-8}$ the current value as 
used by Chopin and Robert \citeyear{chop:rob10}. Fortunately, for the models in question, the full conditionals of the parameters was 
available even for the tempered posteriors. This was exploited by using Gibbs kernels in all of the Monte Carlo experiments.

 

\begin{table}
\begin{center}
\begin{tabular}{l c r}
\hline\hline
Method \qquad & \qquad $\mbox{mean}(BF_{21})$ \qquad& \qquad${\mbox{S.E.}}(BF_{21})$\qquad\\
\hline
Exact & 4553.65 & $-$\\
Laplace approximation & 4553.63 & $-$\\
Laplace approximation MAP & 4553.74 &  1.05 \\
Harmonic mean estimator & 3827.47 & 768.31 \\
Chib's method & 4553.69 & 0.66 \\
Annealed importance sampling &4597.89 & 181.33\\
Nested sampling & 5369.71 & 3874.79\\
Power posteriors & 4556.36 & 66.90\\
\hline
\end{tabular}
\end{center}
\caption{Comparison of different approaches to estimating the Bayes factor of Model 2 over Model 1 based on eighteen runs of 
each algorithm for the Radiata Pine data.} \label{tab:comparison}
\end{table}

In this simple experiment it seems that Chib's method performs very well. Of course this method assumes that the full conditionals 
are readily available and easy to sample from. The poorest performances are from the harmonic mean estimator and nested sampling, 
especially when looking at the standard error of the Bayes factor. Issues with the harmonic mean estimator have been mentioned 
earlier. It is possible that the performance of the nested sampling algorithm could be improved by better simulation from 
constrained distribution at high values of the likelihood, or maybe a stricter termination criterion. With the criterion used 
however, run times could still be time consuming. Chopin and Robert \citeyear{chop:rob10} have proposed nested 
importance sampling, and suggested more efficient ways to simulate from the constrained likelihood distributions. This may 
alleviate some of the issues encountered here. It should be kept in mind that this example is quite straightforward. The simpler 
methods presented here may not work so well for more elaborate models.

The results presented here should be interpreted as a rough guide to accuracy of the various methods. We acknowledge that there 
are various ways in which some of the algorithms could be optimized, for example the tempering scale in annealed importance 
sampling and power posteriors, or the constrained sampling step in the nested sampling algorithm. We have employed by-and-large 
vanilla schemes here. For complete transparency, we have made the code available which was used to obtain these results at 
\texttt{www.ucd.ie/statdept/jwyse/Evidence.zip}, including some documentation explaining the inputs to the code. We invite readers to 
experiment with different specifications to the algorithms and get a feel for the sensitivity of the results to prior assumptions.

\subsection{Choosing between two logistic regression models} \label{ex:logistic}

Here we examine the Pima Indians data which records instances of diabetes and a range of possible diabetes indicators for 
$n = 532$ Pima Indian women aged 21 years or over. There are seven potential predictors of diabetes recorded for this group; 
number of pregnancies (\texttt{NP}); plasma glucose concentration (\texttt{PGC}); diastolic blood pressure (\texttt{BP}); triceps 
skin fold thickness (\texttt{TST}); body mass index (\texttt{BMI}); diabetes pedigree function (\texttt{DP}) and age (\texttt{AGE}). 
This gives 129 potential models (including a model with only a constant term). Diabetes incidence ($y$) is modelled by the likelihood
\[
\pi(y|\theta) = \prod_{i=1}^n p_i^{y_i}(1-p_i)^{1-y_i}
\]
where the probability of incidence for person $i$, $p_i$, is related to the covariates (including constant term) 
$x_i = (1,x_{i1},\dots, x_{id})^{\transpose}$ and the parameters $\theta = (\theta_0,\theta_1,\dots,\theta_d)^{\transpose}$ by
\[
\log\left(\frac{p_i}{1-p_i}\right) = \theta^{\transpose} x_{i}
\]
where $d$ is the number of explanatory variables. An independent multivariate Gaussian prior is assumed for the elements of 
$\theta$, so that
\[
\pi(\theta) =  \left(\frac{\tau}{2\pi}\right)^{d/2} \exp\left\{-\frac{\tau}{2}\theta^{\transpose}\theta \right\}.
\]
The covariates were standardized before analysis. 

A long reversible jump run \cite{gre95} revealed that the two models with the highest posterior probability were
\begin{center}
$\begin{array}{c}
\mbox{Model 1: \texttt{logit(p) $=$ 1 $+$ NP $+$ PGC $+$ BMI $+$ DP }}\\
\mbox{and}\\
\mbox{Model 2: \texttt{logit(p) $=$ 1 $+$ NP $+$ PGC $+$ BMI $+$ DP $+$ AGE }}.
\end{array}$
\end{center}
This reversible jump algorithm assumed a non-informative value of $\tau=0.01$ for the prior on the regression parameters. For this value of $\tau$ we carried out a reduced reversible jump run restricting to jumps only between these two models. The prior probabilities of the models were adjusted to allow for very frequent jumps (about 29\%). This gave a Bayes factor $BF_{12}$ of 13.96 which will be used as a benchmark to compare the other methods to. 

\begin{table}
\begin{tabular}{lcccr}
\hline \hline
Method & $\log \pi(y|\mbox{Model 1})$ & $\log \pi(y|\mbox{Model 2})$ & $BF_{12}$ & Relative speed \\
\hline
Laplace approximation & -257.26 & -259.89 & 13.94 & 1\\
Chib \& Jeliazkov's method & -257.23 & -259.84 & 13.66 & 44\\
Laplace approximation MAP & -257.28 & -259.90 & 13.77 & 108\\
Harmonic mean estimator & -279.47 &-284.78& 203.12 & 108 \\
Power posteriors & -257.98 & -260.59 & 13.71 & 184\\
Annealed importance sampling & -257.87 & -260.43 & 12.83 & 194\\
Nested sampling & -258.82 & -261.38 & $ 12.99 $ & 808\\
\hline
\end{tabular}
\caption{Estimated log marginal likelihoods for each model and corresponding Bayes Factors for each method along with relative 
run times with $\tau = 0.01$.} \label{tab:pima}
\end{table}

We compared each of the methods in Section~\ref{sec:methods} in estimating $BF_{12}$. The results are shown in Table~\ref{tab:pima} 
ranked according to the speed of each of the methods. The results given for each of the methods were the best approximation to 
$BF_{12}$ over five runs of the experiment. It was not possible to carry out an extensive simulation study here due to time 
constraints.  The approximate relative speed is calculated by taking the average execution times of each of the algorithms over 
five runs of the experiment. 

Some measures were taken to try and make the implementations of each scheme as fair as possible for comparison in terms of speed, 
although we do not claim that our implementations are optimal. The code is available from \texttt{www.ucd.ie/statdept/jwyse/Evidence.zip}. Each 
Monte Carlo method used the equivalent of 200,000 samples. For example, the power posteriors use 20,000 samples at each of 10 steps. 
The annealed importance sampling algorithm uses a tempering ladder of 101 to generate 2000 independent samples from the posterior. 
The Laplace at MAP and harmonic mean estimator both use 200,000 samples. Nested sampling uses 2,000 samples and is terminated when 
a new contribution to the approximation to $\pi(y)$ is less than $10^{-8}$ times the current value. 

As there are no full conditionals available for the tempered posteriors in this instance, some care is needed as to how to scale 
proposals within different (inverse) temperatures. It is desired to have wider proposals at lower values of $t$ so that the 
algorithm can freely explore the support of the posterior. For this reason, when updating $\theta_j$ in temperature $t$ we chose a 
proposal from a Gaussian distribution which is centered at $\theta_j$ and with standard deviation $(t^{\alpha}\tau_p)^{-1/2}$. Here 
we chose $\tau_p = 2$ and the value of $\alpha$ is so that the variability of the proposal near zero temperature would equal that 
of the prior. In the case of power posteriors, this was
\[
\alpha = \log(\tau/\tau_p)/\log(t_{1}).
\]


The most notable performances from Table~\ref{tab:pima} are those of the harmonic mean estimator and nested sampling. Something 
which is of note here is the sensitivity of the Bayes factor to the value of $\tau$. For example, increasing $\tau$ to 1, and hence 
being more informative about the variability of parameter values, reduces this Bayes factor to about 1.3 which would lead to a much 
weaker conclusion as to which was the better model for the data. When the experiment is run with $\tau=1$ we note much better 
performance of all the algorithms in estimating $BF_{12}$ (Table~\ref{tab:pima2}). A value of $BF_{12} = 1.3$ was obtained from a 
long reversible jump run. 

\begin{table}
\begin{tabular}{lcccr}
\hline \hline
Method & $\log \pi(y|\mbox{Model 1})$ & $\log \pi(y|\mbox{Model 2})$ & $BF_{12}$ & Relative speed \\
\hline
Laplace approximation & -247.33 & -247.59 & 1.31 & 1\\
Chib \& Jeliazkov's method & -247.31 & -247.58 & 1.32	 & 40\\
Laplace approximation MAP & -247.33 & -247.62 & 1.34	 & 98\\
Harmonic mean estimator & -259.84 & -260.55 & 2.03 &  98\\
Power posteriors & -247.57 & -247.84 &  1.31 & 169\\
Annealed importance sampling & -247.30 & -247.59 & 1.33 & 178\\
Nested sampling & -246.82 & -246.97 & 1.15 & 610\\
\hline
\end{tabular}
\caption{Estimated log marginal likelihoods for each model and corresponding Bayes Factors for each method along with relative run 
times with $\tau = 1$.} \label{tab:pima2}
\end{table}


\section{Concluding thoughts}

The evidence is a fundamental quantity in Bayesian statistics. This short article has surveyed some widely
used approaches in the literature and hopefully has encouraged interested readers to explore these methodologies
further. The evidence is generally a difficult quantity to estimate, especially if the prior distribution is
diffuse, and as such, it is not unsurprising that it may take some effort to implement, in terms of computational 
run time and computer coding. Often, the easiest to implement method may not give the most reliable estimates.

\section*{Acknowledgements}
Nial Friel was supported by a Science Foundation Ireland Research Frontiers Program grant, 09/RFP/MTH2199.

\bibliography{hmrf}

\end{document}